\documentclass[preprint,showpacs,aps]{revtex4}
\usepackage{graphicx,amssymb}
\begin{document}
\draft
\newcommand{\be}{{\,\bf e}}
\newcommand{\bp}{{\bf p}}
\newcommand{\bq}{{\bf q}}
\newcommand{\br}{{\bf r}}
\newcommand{\bv}{{\bf v}}
\newcommand{\bE}{{\bf E}}
\newcommand{\bF}{{\bf F}}
\newcommand{\bH}{{\bf H}}
\newcommand{\bJ}{{\bf J}}
\newcommand{\bM}{{\bf M}}
\newcommand{\dfrac}{\displaystyle\frac}
\newcommand{\tJ}{{\hat{\bf J}_{\rm M}}}

\title{Non conservation of the magnetization current across
magnetic hetero-structures}
\author{S. T. Chui and Z. F. Lin}
\affiliation{ Bartol Research Institute,
University of Delaware, Newark, DE 19716}

\date{\today}

\begin{abstract}
We show that when the magnetizations on opposite sides of a junction are
not collinear, the magnetization current is not conserved as the junction
is crossed. Thus the usual treatment of this problem needs to be modified.
We argue that this is due to an implicit assumption of
an external torque that is required to clamp the magnetization in place.
The physical consequence of this is explored.
\end{abstract}

\pacs{PACS numbers:73.40.-c,71.70.Ej,75.25.+z}
\maketitle There is much recent interest in spin polarized
transport in heterostructures such as tunnel junctions and
metallic multilayers. A general approach to these types of
problems is to solve for the steady state solutions. To
incorporate the transmission and reflection of the injecting
current at the interface, it is necessary to solve the boundary
value problem. Central to treating the transmission and the
reflection is how the charge current $J$  and magnetization current
$J_{\bf M}$ go from one side of the heterostructure to the next.
For a simple heterostructure with no magnetic impurities both
types of currents are usually assumed to be conserved as they go
across the barrier. An equation that is usually employed in the
treatment of this phenomena is of the form\cite{Valet}
\begin{equation}
J_{\bf M}
=\sum_s (\alpha_s\mu_{Ls}-\beta_s\mu_{Rs})
\end{equation}
where $\alpha$, $\beta$ are coefficients, $\mu_{L(R)s}$ are the
chemical potentials of the electrons of spin channel $s$ on the left
(right) of the junction. In this equation, it is assumed that
$J_{\bf M}$ is unique because it is conserved. Recently, it was
discovered that when the magnetizations on opposite sides of a
metallic barrier structure are noncollinear, a current can induce
torques on the magnetizations\cite{Cornell,Tsoi,Slon1,Berger}.
This motivated us to examine the boundary condition across a
heterostructure more closely. We consider the strong barrier limit
corresponding to tunnel junctions and the weak barrier limit with
zero interface resistance, corresponding to metallic multilayers.
We find that when the magnetizations on opposite sides of the
barrier are noncollinear, the magnetization current is {\bf not}
conserved as it goes across the barrier. Thus eq. (1) cannot
always be used. An examination of the physics suggests that {\bf
implicit} in the assumption of the noncollinear magnetizations is
a torque to maintain the orientation of the magnetization. It is
this torque which changes the magnetization current. We now
discuss our results in detail.

We first remind the reader the meaning of the direction of
magnetization. The general quantum statistical properties of a
system is described by the density matrix $\rho$. The general
state of magnetization is therefore described by the $2\times2$
polarization matrix $|M_0|\rho$ where $|M_0|$ is the magnitude of
the magnetization. As is well known\cite{Merzbacher} there is
always a direction ${\bf M}$ so that $\rho=(1+{\bf \sigma \cdot
M})$ where ${\bf \sigma}$ are the Pauli matrices
$(\sigma_x,\sigma_y,\sigma_z)$. The density matrix is diagonal
along ${\bf M}$. We call this the direction of magnetization.
Along this direction, the off-diagonal components of the density
matrix are zero and there is no phase coherence between the spin
up and the spin down states.

We next evaluate the magnetization current. For simplicity we take a
one dimensional geometry where the currents are uniform in the $x$-$z$
plane and
move in the $y$ direction. The evaluation of the magnetization current
is simplest
when we pick a coordinate system along the direction of magnetization.
The magnetization current
then becomes the difference between the spin up and the
spin down current. We first consider the strong barrier limit of
spin polarized tunnelling. Our starting point is the well accepted tunnelling
Hamiltonian $H_T=\sum T_{ss'}c^+_{Lks}c_{Rqs'}+c.c.$
where $s$, $s'$ are with respect to the magnetization directions on the left
(labelled by the subscript $L$)
and on the right (labelled by the subscript $R$). $k$, $q$ are other labels
for the electronic states.
We are interested in the case when the directions of
magnetization on opposite sides of the junction are not the same.
The spins states $|s\rangle$ along one magnetization direction are related
to those ($|s'\rangle $) along the other direction by a rotation matrix ${\bf R}_{s,s'}$.
 From a single particle quantum mechanical
calculation we obtain an estimate of the tunnelling
matrix element
\begin{equation}
T_{ss'}={\bf R}_{ss'}\exp[\int_0^d dy
(\hbar^2(U-\mu_{Ls}+\mu_{Rs'}-Vy)/2m)^{0.5}].
\end{equation}
where $U$ is the tunnelling barrier, $V$ is the external voltage, $m$
is the electronic mass. In linear response, the term
$-\mu_{Ls}+\mu_{Rs'}-Vy$ is absent.

Stimulated by the experimental success in superconductivity
tunnelling there has been much studies done with this type of
tunnelling Hamiltonian. We shall follow the work by Cohen, Falicov
and Phillips\cite{CFP}, as is explained in detail in
Mahan\cite{Mahan}. The magnetization current on the left can be
evaluated from the time rate of change of the left magnetization
due to the tunnelling Hamiltonian: Since $J_s=-\partial
N_s/\partial t$, we get $J_M^L=-\partial (N_+^L-N_-^L)/\partial
t=- \partial M_L/\partial t$ (we have used a unit so that the Bohr
magneton is 1) with a
similar expression for $J_M^R$.  
The time derivative of the particle number of spin $s$ can be evaluated
as its commutator with the tunnelling Hamiltonian. The details of this for
each spin component is essentially the same as in previous
calculations and thus will not be repeated here.  We find that\cite{cf}
\begin{equation}
J_M^L=4\pi e \sum_{ss'}s|T{ss'}|^2 (n_L(\mu_{Ls})-n_R(\mu_{Rs'})),
\end{equation}
and
\begin{equation}
J_M^R=4\pi e \sum_{ss'}s'|T{ss'}|^2 (n_L(\mu_{Ls})-n_R(\mu_{Rs'})).
\end{equation}
where $\mu_{Ls}$ ($\mu_{Rs'}$)
is the chemical potential for the spin $s$ ($s'$) electrons on the
left (right) side. $n(\mu)$ is the density of particles at energy $\mu$.

When the directions of magnetization on both sides of the junction are
collinear, $T$ is diagonal and we find that
$J_M^L=4\pi e \sum_{s}s|T{ss}|^2
(n_L(\mu_{Ls})-n_R(\mu_{Rs}))=J_M^R.$ The magnetization current is conserved.
When the magnetization directions are not collinear, $T$
is no longer diagonal.
In the linear response regime, $T_{s,s'}=T_{s',s}$ is a symmetric
matrix. After renaming the variables of summation, one has
$J_M^L=4\pi e \sum_{ss'}s'|T{ss'}|^2 (n_L(\mu_{Ls'})-n_R(\mu_{Rs}))$.
$J_M^L\neq J_M^R$ because $\mu_{Ls'}\neq \mu_{Ls}$ for $s\neq s'$.
The latter comes about because there is a splitting of the spin
up and spin down chemical potential caused by the spin accumulation
effect\cite{6,7} and the interplay between the charge and magnetization
dipole layers\cite{1}.

$J_M^L$ and $J_M^R$ correspond to magnetizations along different
directions. In the above we have shown that the magnitude of these
two currents are not the same. A more appropriate condition would
be requiring the magnetization currents for magnetizations along
the same direction be equal: $J_M^L=J_M^R$. This
requirement obviously cannot be satisfied because for example, on
the right hand side it would require a magnetization current with
a magnetization component that is perpendicular to the
magnetization on the same side. From what we just derive, this is
zero. The reason for the nonconservation of the magnetization
current is easy to understand. Consider a wave function $|sL\rangle$ of
an electron of spin $s$ on the left. With respect to the
magnetization direction on the right, this wavefunction can be
written as $\sum_{s'}{\bf R}_{ss'}|s'R\rangle$. The corresponding
density matrix for a state like this is given by
$$
\rho_0=\left (
\begin{array}{cc}
R_{++}^2 & R_{+-}^2\\
R_{+-}^2 & R_{--}^2
\end{array}
\right )
$$
with non-zero off-diagonal elements. But by our definition of the
magnetization direction, we have assumed that after this electron
has moved from the left to the right, the phase coherence between
the up and down states are lost and the off-diagonal components
become zero. This happens because we have {\bf implicitly} assumed
that there is some external torque that clamps the magnetization
in certain directions. This torque can come from the crystalline
anisotropy energy of the system, for example. It is this implicit
torque that causes the magnetization current not to be conserved.
In addition to this torque $t_0=(J_M^R-J_M^L)/a\mu_B$ ($\mu_B$ is
the Bohr magneton, $a$ is atomic spacing) there may be other
contributions which we shall not discuss in this paper. The
reasoning in this section also applies to metallic multilayers. We
turn our attention to that case next.

For metallic multilayers, eq. (3) and (4) are no longer valid.
For simplicity we consider the other limit where the interface resistance
is zero. For that case, the current is controlled by the voltage drop
along the metal. We assume that the length of heterostructure is less
than the spin diffusion length and apply a parallel circuit model to
illustrate the essential physics.
In that case, we get $J_s^{L(R)}=\sigma_s^{L(R)}E^{L(R)}$
where $\sigma_s^{L(R)}$ is the conductance for spin $s$ on the left (right).
$E$ is the electric field that drives the current. The spin current on opposite
sides of the heterostructure are related by
 $J_{s'}^L=\sum_s|{\bf R}_{s,s'}|^2J_s^R$,
where, again, ${\bf R}$ is the rotation matrix. There are two ways that
one can see this. The current on the right for
spin index $s'$ is just the number of electrons with $z$ angular momentum
$s'$ moving past per unit time
per unit area. These electrons come from the left side.
The factor $R^2$ takes into account the probability that the $z$
angular momentum is $s'$.  Formally,
in terms of electron creation and destruction operator
the current operator is just $J_{s'}^R=\langle\sum v c_{s'}^+c_{s'}\rangle$
where $v$ is the velocity.
Now the creation and destruction operator can be expressed with the z axis
along $M_L$ as $c_{s'}={\bf R}_{s',s}c_s$, $c_{s'}^+={\bf R}_{s',s}c_s^+.$
Because ${\bf R}$ is real, it occurs in the expressions for both
the creation and destruction operator. By our assumption of the
magnetization direction, we have set such terms as $<\sum c_s^+c_t>$
with $s\neq t$ to be zero. Hence the final expressions quoted above is
obtained.

It is straightforward
to show that $E^{L(R)}=\sigma^{L(R)}E/(\sigma^L+\sigma^R)$ where
the total conductance $\sigma^{L(R)}=\sum_s\sigma_s^{L(R)}$,
$E=E^L+E^R$ measures the total voltage drop. Just as explained in the previous
paragraph, because the magnetization points in different directions on
opposite sides of the heterostructure, the magnetization currents are not
equal to each other. The magnitudes of the two currents are equal to each other,
however.

In the presence of a finite interface resistance, in linear response
the current is linearly proportional to the chemical potential. We thus expect
the physics to be governed by equations of the form
\begin{equation}
J^L_M=\sum_s r_{1s}^L\delta \mu_s^L-r_{2s}^L\delta \mu_s^R
\end{equation}
\begin{equation}
J^R_M=\sum_s r_{1s}^R\delta \mu_s^L-r_{2s}^R\delta \mu_s^R
\end{equation}
where the $r$'s are the effective interface resistances.
We next explore the physical implication of our result.

In typical experimental arrangements, only the directions of the magnetizations
{\bf far away} from the interface are specified. Because of the torque, these
directions will be rotated as the interface is approached. A
domain wall is created. We estimate
this rotation here for a simple case. We assume that far from the interface
the magnetzations are in the $x$-$z$ plane along the easy axis of anisotropy
at angles $\theta_L$, $\theta_R$ so that $\theta_R-\theta_L=\Phi$.
At the interface, the left (right) magnetization has rotated by an
angle $-\delta \theta_R$ ($\delta \theta_L$). From minimizing the energy,
the equation governing the angle $\delta \theta$ at a distance $y$ away
from the interface is given by $J_i d^2\delta \theta/dy^2-K_i\delta \theta
=0$ where $i=L, R$; $J_i$ ,$K_i$ are the exchange and anisotropy constants.
The effective field $H_{eff}$ due to the exchange, $
J_i d^2\delta \theta/dy^2$, is balanced out by that due to the
anisotropy, $K_i\delta \theta$.
The solution of this equation is $\delta\theta (y)=\delta\theta_i
exp(-|y|/d_i)$ where the magnetic length $d_i=\sqrt{J_i/K_i}$.
At the interface, the effective field from the exchange
becomes $Jd\delta\theta/dy$ while that from the anisotropy
remain the same. The net torque
$t=-\partial E/\partial\theta=M_0\delta\theta_i(K_i- J_i/d_i)
=M_0\delta\theta_i[K_i- (J_iK_i)^{0.5}]$.
This creates the difference in the magnetization current discussed above.
We thus get $t=t_0$.
$\delta\theta_i
=\Delta J_M/M_0a[K_i- (J_iK_i)^{0.5}]$.
Thus we can relate $\delta\theta_i$ to $\Delta J_M$.

Experimental measurements are carried out under constant charge
current conditions. It is necessary to relate $\Delta J_M$ to the
current $J$. To achieve this goal it is necessary to find
solutions for the currents for each side of the junction and match
them across the junction. In the treatment of spin polarized
transport, we should use eq. (3) and (4) instead of eq. (1). These
equations will be complemented by equations which relate the
currents to the chemical potentials inside the material on either
side of the junction. An example of such a equation is the
modified Landau Gilbert equation with a source term equal to the
divergence of the magnetization current:\cite{EB}
\begin{equation}
\frac{\partial \bM }{\partial t} - \gamma \bM\times\bH +
 \nabla\cdot\tJ = -\frac{\delta\!\bM}{\tau}
\end{equation}
where $\gamma$ is the gyromagnetic ratio, $\tJ$ is the spin current
(tensor), $\tau$ is the spin relaxation time,
and $\bH$ is the effective field
describing the precession of the magnetic moments.
The magnetization and charge currents are given by Fick's law as sums of terms
proportional to the gradients of the charge ($\delta n$) and magnetization
($\delta {\bf M}$) densities and the
local internal electric field. They are
\begin{eqnarray}
\bJ & = &
-\sigma\nabla V
      -\beta D \nabla \delta\!\bM \cdot \bp_0
      - \alpha_1 D \nabla \delta n  \\
\tJ & = &
-\beta\sigma \nabla V \bp_0 -D\nabla\delta\!\bM
-\alpha_2 \beta D \nabla \delta n \bp_0
\end{eqnarray}
where $\alpha_1$ and $\alpha_2$ are two phenomelogical parameters,
$\beta$ measures the asymmetry of the spin up and spin down current
in the metal, $\sigma$ is the conductivity of metal, and $D$ is the
diffusion coefficient. ${\bf p}_0$ is the unit vector along the
direction of magnetization.
$V$ is the sum of the electic potential $V_e$
describing the external electric field and the local electric
(screening) potential $W$ due to other electric charges determined
self-consistetly
$$
W({\bf r})
=\int d^3{\bf r}' U({\bf r}-{\bf r}') \delta n({\bf r}').
$$
From these equations and the charge conservation
\begin{equation}
\nabla\cdot{\bf J}=-\frac{\partial \delta n}{\partial t}
\end{equation}
we obtain
for each side of the heterostructure
the charge and magnetization dipole layers at the interface given
mathematically by
\begin{equation}\begin{array}{ll}
\delta\! n^R =  \displaystyle \sum_{i=1}^{4}
         \delta\! n^R _{i0}e^{-(y-\frac{d}{2})/l_i}, \\
\delta\! \bM^R  =\displaystyle \sum_{i=1}^{4}
        \delta\! \bM^R _{i0}e^{-(y-\frac{d}{2})/l_i}
\end{array}
\end{equation}
where the four length scales correspond to the screening length and the
magnetic lengths for the longitudinal and left and the right rotating
transverse magnetizations. $d$ is the thickness of the barrier.
The superscript $R$ denotes the right hand side.
A similar expression can be written for quantities
on the left hand side.  Because of the coupling between the charge and
spin degrees of freedom, the length scales $l_i$ are renormalized
from their bare value.
The coefficients $\delta n_{i0}$ are found to depend on
$\delta {\bf M}_{i0}$,
while $\delta {\bf M}_{i0}$ are
determined by matching the boundary condition at the interface.
We illustrate this approach with a calculation of the steady state solutions
in the low field linear response limit.
After this boundary matching is carried out, we found for
metallic multilayers in linear response
a large change in the {\bf longitudinal} component of the magnetization that
is of the order of the voltage times the density of states at the Fermi
surface (in units with $\mu_B=1$). The transverse magnetization change
is much smaller. As a result, there is no additional torque generated
along the ferromagnet. All the torque comes at the interface.

When the material parameters on both sides are the same, we
found
\begin{equation}\begin{array}{ll}
e\Delta J_M =  \beta J \cos(\Phi/2)\sin\Phi \\
\times \dfrac{A_1 +2\sqrt{1-\beta^2} \xi \bigl[ 1
       +A_2\sin^2(\Phi/2)/(1-\beta^2)\bigr]
       +\xi^2 4\sin^2\Phi/2 } {1-\beta^2
       +2\sqrt{1-\beta^2} \xi +\xi^2\sin^2 \Phi}
\end{array}
\end{equation}
where $\xi$ is the ratio of the conductance of the interface
to that of the metal.
$A_1$, $A_2$ are some dimensionless
constants of the order of unity.

In summary, we argue that the magnetization current is not conserved
at the interface of a heterostructure when the magnetizations on
opposite sides of the structure are not collinear. This is due to the
implicit assumption of a torque which creates a domain wall at the interface.
This is illustrated with a steady state solution of the of the problem
in the limit of linear response. In that limit, all the torque
comes at the interface, no additional torque is introduced in the metal.

Z L was supported in part by China NNSF.


\begin{references}
\bibitem{cf} This is basically eq. (9.39) in \cite{Mahan}.
\end{references}
\end{document}